\documentstyle[preprint,aps]{revtex} 
\tightenlines

\begin{document}

\title{Riemann-Einstein Structure from Volume and Gauge Symmetry}

\author{Frank Wilczek}
\address{Institute for Advanced Study, School of Natural Science, Olden Lane, 
Princeton, NJ 08540}
%\date{January 20, 1998}
\preprint{IASSNS-HEP-97/142}
\maketitle

\begin{abstract}
It is shown how a metric structure can be induced in a simple way
starting with a gauge structure and a preferred volume, by spontaneous
symmetry breaking.  A polynomial action, including coupling to matter,
is constructed for the symmetric phase.  It is argued that assuming a
preferred volume, in the context of a metric theory, induces only a
limited modification of the theory.
\end{abstract}

\bigskip
\S1 Attempts to Use a Gauge Principle for Gravity 
\bigskip

The fundamental laws of physics, as presently understood, largely
follow from the two powerful symmetry principles: general covariance,
and gauge invariance.  There have been many attempts to unite these
two principles.  Indeed the very term, gauge invariance, originates
from an early attempt~\cite{weyl} by Weyl to derive classical
electromagnetism as a consequence of space-time symmetry, specifically
symmetry under local changes of length scale.  The modern
understanding of gauge invariance, as a symmetry under transformations
of quantum-mechanical wave functions, was reached by Weyl himself and
also by London very shortly after the new quantum mechanics was first
proposed.  In this understanding of abelian gauge invariance, and in
its nonabelian generalization~\cite{yang}, the space-time aspect is
lost.  The gauge transformations act only on internal variables.  This
formulation has had great practical success. Still, it is not entirely
satisfactory to have two closely related, yet definitely distinct,
fundamental principles, and several physicists have proposed ways to
unite them.

One line of thought, beginning with Kaluza~\cite{kaluza} and
Klein~\cite{klein}, seeks to submerge gauge symmetry into general
covariance.  Its leading idea is that gauge symmetry arises as a
reflection in the four familiar macroscopic space-time dimensions of
general covariance in a larger number of dimensions, several of which
are postulated to be small, presumably for dynamical reasons.  Here we
should take the opportunity to emphasize a point that is somewhat
confused by the historically standard usages, but which it is vital to
have clear for what follows.  When physicists refer to general
covariance, they usually mean the form-invariance of physical laws
under coordinate transformations following the usual laws of tensor
calculus, including the transformation of a given, preferred metric
tensor.  Without a metric tensor, one cannot form an action principle
in the normal way, nor in particular formulate the accepted
fundamental laws of physics, viz. general relativity and the Standard
Model.  From a purely mathematical point of view one might consider
doing without the metric tensor; in that case general covariance
becomes essentially the same concept as topological invariance.  The
existence of a metric tensor reduces the genuine symmetry to a much
smaller one, in which space-times are required not merely to be
topologically the same, but congruent (isometric), in order to be
considered equivalent.  In the Kaluza-Klein construction, for this
reason, the gauge symmetries arise only from {\it isometries\/} of the
compactified dimensions.

Another line of thought proceeds in the opposite direction, seeking to
realize general covariance -- in the metric sense -- as a gauge
symmetry.  A formal parallelism is readily perceived~\cite{utiyama}.
Indeed, it is quite helpful, in bringing spinors to curved space, to
introduce a local gauge $SO(3,1)$ symmetry \footnote{For definiteness
I consider 3+1 dimensional space here.}.  The spin connection
$\omega_\mu^{ab}$ is the gauge potential of this symmetry.  The lower
index is a space-time index, and the upper indices are internal
$SO(3,1)$ indices, in which $\omega$ is antisymmetric, as is
appropriate for the gauge potential.  In this construction one must,
however, introduce an additional element that is not a feature of
conventional gauge theories.  That is, one postulates in addition the
existence of a set of vierbein fields $e^a_\alpha$ with the properties
\begin{equation}
\eta_{ab} e^a_\alpha e^b_\beta =
g_{\alpha \beta} 
\label{eq:mfrome}
\end{equation}
\begin{equation}
g^{\alpha \beta} e^a_\alpha e^b_\beta = \eta^{ab}
\label{eq:etafrome}
\end{equation}
where $\eta$ is the internal space
metric tensor.  The $e^a_\alpha$ are supposed to transform
homogeneously under the local symmetry, and to behave as ordinary
vectors under general coordinate transformations.  The $e^a_\alpha$
link the internal and space-time indices, so that by contracting with
them one can freely trade one kind of index for the other.  

The formal nature of this local $SO(3,1)$ symmetry is evident from
(\ref{eq:mfrome}).  The action of the symmetry simply interpolates
between the different possible choices for the solution of this
equation, all corresponding to the same space-time metric.  Thus given
the metric the vierbein adds no gauge-invariant information, at least
locally.  One could equally well have both a vierbein field and the
gauge symmetry which effectively says it is arbitrary, or neither.
For coupling to spinors, of course, it is very convenient to make the
former choice.  Another difference between the formal $SO(3,1)$ gauge
invariance of general relativity and the usual gauge symmetries of the
Standard Model is the postulated action.  In general relativity the
Einstein-Hilbert action density, in terms of the $SO(3,1)$ gauge field
strength and the vierbeins, is given as
\begin{equation}
{\cal L} ~=~
\det e~ e^\alpha_a e^\beta_b F^{ab}_{\alpha \beta}
\label{eq:EHaction}
\end{equation}
where the determinant involves
$e^a_\alpha$ regarded as a square matrix, and the $e^\alpha_a$ are
elements of the inverse matrix.  The existence of an inverse to the
introduced $e^a_\alpha$ is essential for expressing the action in this
form.  In this action the $\omega$ appear, after an integration by
parts, only algebraically.  They can therefore be eliminated in favor
of the vierbeins, or ultimately the metric.  All this is quite unlike
what we have for conventional Yang-Mills theory.  

A markedly different approach to casting gravity as a gauge theory was
initiated by MacDowell and Mansouri~\cite{macdowell}, and
independently Chamseddine and West~\cite{west}.  It has proved
useful as a way of introducing supergravity.  Their construction, as
improved by Stelle and West~\cite{stelle}, is along the following lines.
One introduces an $SO(3,2)$ gauge symmetry (or $SO(1,4)$; $SO(3,2)$ is
more appropriate for supergravity), with the associated gauge
potentials $A^{ab}_\mu$.  One introduces as well a space-time scalar,
internal-space vector field $\phi^a$.  Very important: no vierbein or
metric is introduced.  The action density is taken to be
\begin{equation}
{\cal L}_{\rm MM} ~=~ \epsilon_{abcde} \epsilon^{\alpha \beta
\gamma \delta} F^{ab}_{\alpha \beta} F^{cd}_{\gamma \delta} \phi^e
\label{eq:MMaction}
\end{equation}
and $\phi^e$ is taken to be frozen at
$\phi^e = \kappa \delta^e_5$.  Here of course the $\epsilon$s are
numerical totally antisymmetric symbols, and $\kappa$ is a constant.
At this point one must distinguish two classes of gauge potentials,
the $A^{a5}_\alpha \equiv h^a_\alpha$ and the $A^{ab}_\alpha$ with
neither $a$ nor $b$ equal to 5.  The field strengths $F^{a5}$ do not
appear in (\ref{eq:MMaction}).  The $h^{a}_{\alpha}$ appear in the remaining
field strengths $F^{ab}$ only in commutator terms.  Let us separate the
terms which are independent of, linear, or quadratic in these
commutators.  The term  independent of the $h^a_\alpha$ is
quadratic in the conventional $SO(3,1)$ potentials and is a total
divergence. It does not contribute to the classical equations of
motion.
The term which is linear in the commutators  -- 
and therefore also linear in the conventional $SO(3,1)$ field
strength -- is most remarkable.  Indeed, if we assume that the matrix
$h^a_\alpha$ is non-singular, and make the identifications
\begin{equation}
\omega_\mu^{ab} \equiv A_\mu^{ab}
\label{eq:spingauge}
\end{equation}
\begin{equation}e^a_\alpha \equiv
h^a_\alpha = A^{a5}_\alpha \label{eq:viergauge}
\end{equation}
we find
that this term is proportional to the Einstein-Hilbert action density.
Finally, in this interpretation, the
term quadratic in the commutators yields a cosmological constant.  

This
construction certainly makes a more compelling connection of gravity
to gauge theory than the preceding formal construction, 
and has the elegance of simplicity.  Yet it is
not entirely satisfactory, for several reasons.  The assumed constancy
of $\phi^e$ is quite {\it ad hoc}, and amounts to a way of sneaking in
some assumptions about a higher symmetry, $SO(3,2)$ or $SO(1,4)$, that
is not really present in the theory.  Likewise, the expansion
around non-singular $h^a_\alpha$ seems arbitrary, especially since
these fields do not transform homogeneously under the higher symmetry
(they do transform homogeneously under the residual $SO(3,1)$).
Perhaps its most important limitation, however, is that it does not
extend in any obvious way to include couplings to matter.  In the
remainder of this note I propose a way to remove these
difficulties.

\bigskip
\S2 Importance of the Volume Element for Index
Contraction 
\bigskip

Concretely, the difficulty with coupling to matter is as
follows.  To obtain a meaningful dynamics, one must include
derivatives for the matter fields.  These are all lower, or covariant,
indices.  In the absence of a metric, it is very difficult to soak up
these indices and form an invariant action.  The only contravariant
object that is intrinsically defined is the numerical antisymmetric
symbol $\epsilon^{\alpha\beta\gamma\delta}$, and it is a tensor
density.  Thus -- assuming full general covariance -- it can be used
to the first power only, to soak up exactly four covariant indices.
This, of course, is how it functioned in (\ref{eq:MMaction}).  In
general, however, and in particular to couple to the fields of the
Standard Model, one seems to require a more flexible way of forming
invariants.  Another aspect of the difficulty is that one cannot form
fully covariant derivative-independent terms, such as are used to
generate masses or effective potentials. With this in mind, it seems
that if one wants to stop short of introducing a metric explicitly, a
natural possibility is to forego full general covariance, and to
require invariance only under transformations that leave the volume
fixed.  Then one can introduce potential terms, and have them introduce
symmetry breaking dynamically, which opens up much wider possibilities
for obtaining realistic theories.
 
\bigskip
\S3 Covariance With a Preferred
Volume Element 
\bigskip

It might seem that relaxing the requirement of full
general covariance would be a very drastic step, and that after taking
it one would have great difficulties in recovering the main
consequences of general relativity.  This is not the case, however, as
I shall now discuss.  If one takes a conventional (metric) general
covariant theory, and considers only variations of the metric that
preserve the volume, one obtains in place of the full Einstein
equations
\begin{equation}R^\mu_\nu - {1\over 2}g^\mu_\nu R ~=~
T^\mu_\nu \label{eq:fullE}
\end{equation}
only their traceless
part
\begin{equation}
R^\mu_\nu - {1\over 4} g^\mu_\nu R ~=~ T^\mu_\nu -
{1\over 4} g^\mu_\nu T~. 
\label{eq:trlessE}
\end{equation}

It might
appear that a great deal has been lost, or modified, in passing from
(\ref{eq:fullE}) to (\ref{eq:trlessE}), but that appearance is
deceptive.  Indeed, (\ref{eq:fullE}) has been written in a way that the
contracted covariant derivative of each side vanishes separately.  For
the left-hand side, this follows from the Bianchi identity.  On the
right-hand side, it follows from {\it full\/} general covariance, with
$T^\mu_\nu$ defined according to the equation 
\begin{equation}
{\delta
S_M \over \delta g_{\mu \nu} } = \sqrt g T^{\mu\nu}
\end{equation}
for variation of the matter action.  We define the
energy-momentum tensor in (\ref{eq:trlessE})  likewise, even
though in the context of that equation we are not requiring that the
action is stationary under full general covariance, after supplying
appropriate powers of $\sqrt g$ to restore full covariance.  Then it too has
zero contracted covariant derivative (i.e., it is covariantly
conserved).  After subtracting $-{1\over 4}g^\mu_\nu R$ from both
sides of (\ref{eq:trlessE}) we can take the contracted covariant
derivative and cancel off the vanishing terms, to obtain
simply
\begin{equation}
\partial_\nu (R + T ) ~=~ 0.
\end{equation}
This
means that $R+T$ is a world-constant -- a number, not a function.
This number is not determined by the gravitational equations of motion,
but is an integration constant, or is determined by other equations of
motion.  In any case, we find that (\ref{eq:trlessE}) differs
from (\ref{eq:fullE}) only in that the cosmological term is no longer
determined by the action and the equations of motion for the metric
tensor, but is an integration constant (perhaps constrained by other
equations of motion).  

We have argued that solutions of
(\ref{eq:trlessE}) yield solutions of (\ref{eq:fullE}) for some value of
the cosmological term, itself undetermined by (\ref{eq:trlessE}).
Conversely, if we suppose that (\ref{eq:fullE}) is valid, but that the
energy-momentum tensor takes the form
\begin{equation}
T^\mu_\nu ~=~T^\mu_\nu (0) + \lambda g^\mu_\nu.
\end{equation}
with $\lambda$
indeterminate, then we can eliminate $\lambda$ by passing to
(\ref{eq:trlessE}). 

It might seem surprising, at first sight, that the
variation of an entire function determines only one number.  However,
of course, general covariance requires that the action is of a very
special form, and indeed invariant under functional transformations,
namely those of general covariance.  Indeed, the situation above could,
from another point of view, easily have been anticipated.  One could,
in the context of orthodox general relativity, take $\sqrt g = {\rm
const.}$ (or any fixed function) as a gauge choice.  The equations of
motion in this fixed gauge must then give the full equations of the
theory, except insofar as fixing the gauge constrains gauge-invariant
quantities.  But the only gauge-invariant quantity associated with
$\sqrt g$ is a single number, the world-volume.  To demonstrate this,
one must show the existence of a general coordinate transformation
having an arbitrary specified Jacobian; but this is, apart from global
questions, evident by counting parameters.  

Other discussions of metric covariance with a fixed volume include 
\cite{zee}.

\bigskip
\S4 Metric Structure from
Symmetry Breaking 
\bigskip

After these preliminaries, the construction of a
metric theory from spontaneous breakdown of gauge symmetry is not
difficult.  The necessary field content is the same as before: an
$SO(3,2)$ gauge field $A^{ab}_\alpha$, and a spacetime
scalar, internal space vector $\phi^a$.  The desired pattern of symmetry
breaking is dynamically favored by action terms of two types.  The first
term we need is of the type
\begin{equation}
{\cal L}_1 ~=~ -\kappa_1
(\eta_{ab} \phi^a \phi^b -
v^2)^2.
 \label{eq:potterm}
\end{equation}
Clearly if $\kappa_1 >0$ this will be stationarized,
and $-{\cal L}_1$ minimized, when $\phi^a (x) = v\delta^a_5$.  Other
solutions can be put in this form by a gauge transformation.  As
$\kappa_1\rightarrow\infty$ this value is effectively frozen in. 

The
second term we need is of the type
\begin{equation}
{\cal L}_2 ~=~
\kappa_2 (J - w)^2
\label{eq:detterm}
\end{equation}
where
\begin{equation}
J ~\equiv~
\epsilon^{\alpha\beta\gamma\delta}\epsilon_{abcde} \phi^e
\nabla_\alpha \phi^a \nabla_\beta \phi^b \nabla_\gamma \phi^c
\nabla_\delta \phi^d 
\label{eq:Jeqn}~,
\end{equation}
and $\nabla$
denotes the  $SO(3,2)$ gauge covariant derivative.  
This term is evidently stationarized at
$J=w$.  If we suppose that $\phi$ takes the form discussed in the
preceding paragraph, then we find
\begin{equation}
J~=~ v^5
\epsilon^{\alpha\beta\gamma\delta} \epsilon_{abcd} A^{a5}_\alpha
A^{b5}_\beta A^{c5}_\gamma A^{d5}_\delta 
\label{eq:deteqn}
\end{equation}
where now the internal space variables run from 1 to 4.  By
stationarizing these two terms, we have reproduced two crucial
elements of the structure described in \S1, that is, an
appropriate symmetry-breaking ``director'' $\delta^a_5$ and a
non-singular ``vierbein'' $e^a_\alpha \propto A^{a5}_\alpha$, as in
(\ref{eq:viergauge}), but now on a dynamical basis, by spontaneously
breaking a legitimate symmetry.  The terms we have used are among the
simplest ones consistent with the assumed symmetries, that is, $SO(3,2)$
or $SO(1,4)$ and volume-preserving reparametrizations.

\bigskip
\S5 Curvature Term 
\bigskip

To complete the identification we should discuss the curvature
term.  In the new framework we can use a simpler construction than
(\ref{eq:MMaction}).  The relevant term is
\begin{equation}
{\cal L}_3 ~=~
\kappa_3 \epsilon^{\alpha\beta\gamma\delta} \epsilon_{abcde}
F^{ab}_{\alpha \beta} \nabla_\gamma \phi^c \nabla_\delta \phi^d \phi^e
~.
\label{eq:newcurv}
\end{equation}
When this is expanded, subject to
the freezing of $\phi$ as before and of the determinant as 
in (\ref{eq:deteqn}),
and with the now familiar identifications, it yields the
Einstein-Hilbert action.  In this form, generalization to any number
of dimensions is immediate.
  
\bigskip
\S6 Matter Couplings 
\bigskip

The structure here
proposed allows coupling to matter fields with no substantial
difficulty, but one subtlety.  Consider, for example, the problem of
constructing a kinetic energy term for a space-time scalar, gauge
scalar field $\psi$.  The standard construction uses both the
determinant of the metric tensor and the inverse metric tensor.  Since
we do not require full general covariance, but only covariance under
volume-preserving reparametrizations, we do not need the determinant.
However we cannot take the inverse metric for granted, for our
effective metric only emerges after ${\cal L}_1$ is stationarized, and
only becomes non-singular when ${\cal L}_2$ is stationarized.  The
appropriate construction is most easily expressed in terms of the auxiliary
field
\begin{equation}
w^\alpha_a ~=~ \epsilon^{\alpha \beta \gamma
\delta} \epsilon_{abcde} \nabla_\beta \phi^b\nabla_\gamma \phi^c
\nabla_\delta \phi^d \phi^e~.  \label{eq:geninverse}
\end{equation}
This
field becomes proportional to the inverse vierbein after our two
steps of symmetry breaking, but is a perfectly well-defined tensor with
the same index structure even in the unbroken phase.  In terms of it,
we can construct the desired kinetic term as
\begin{equation}
{\cal L}_{\rm kin.} ~=~ w^\alpha_a w^\beta_b \eta^{ab} \partial_\alpha \psi
\partial_\beta \psi~.
\end{equation}

One should note that these terms
involve high powers of the covariant derivatives of
$\phi$.

In a very similar fashion, one can construct kinetic terms for
space-time spinor fields.  In this case only one power of $w$ is required.
Yang-Mills fields require the form
\begin{equation}
{\cal L}_{\rm Y-M} ~=~ G^I_{\alpha\beta} G^I_{\gamma\delta} \eta^{ac} 
w^\alpha_a w^\gamma_c \eta^{bd} w^\beta_b
w^\delta_d \label{eq:ymform}
\end{equation}
or more economically
\begin{equation}
{\cal L}_{\rm Y-M}^\prime ~=~  G^I_{\alpha\beta} G^I_{\gamma\delta} 
\epsilon^{\alpha\beta\mu\nu} \epsilon^{\gamma\delta\rho\tau} 
\nabla_\mu \phi^a \nabla_\nu \phi^b \nabla_\rho \phi^c \nabla_\tau \phi^d 
\eta_{ac} \eta_{bd}.
\end{equation}

It is very important, from the point of view of general dynamics, that
the interactions we need are of no higher than second order in time
derivatives.  Thus they have a well posed initial value problem.  This
requirement greatly constrains the form of possible additional terms
\cite{sundrum}.

%%%%%%%%%%%%%%%% 
\bigskip
\S7 On the Cosmological Term
\bigskip

A most puzzling feature of current physical theory is that the vacuum
is supposed to contain several symmetry-breaking condensates
(e.g., electroweak $SU(2) \times U(1)$ breaking and QCD chiral $SU(2)
\times SU(2)$ breaking), yet for purposes of gravity displays very
small or vanishing energy density.  This is the famous problem of the
cosmological term.

In the framework discussed here, this problem takes a very different
appearance.   We have already touched on this in a general way in
\S3.  Now we can illustrate concretely how extra (non-metric) field
equations can augment that discussion.

For present purposes it is useful to introduce symbols $\sigma$ and
$A$ for the vacuum expectation 
values of $\phi^5$ and of $(\det A^{5a}_\alpha )^{1\over
4}$ -- the ``determinantal'' part of the vierbein -- respectively.
Note that one can swing the expectation value 
$\phi$ to lie in a fixed direction, say 5, by appropriate gauge
transformations.

Let us assume that $\sigma$ and $A$ are constants, and examine the
dependence of various possible terms in the 
Lagrangian on these constants.
We have for the `gravitational' terms

\( \begin{array}{clrr}
\lambda_{1} (\phi^{a}\phi^{b} \eta_{ab})^{2} &\sim \lambda_{1}
\sigma^{4} &;  [\lambda_{1}] = &4 \\

\lambda_{2} (\phi^{a}\phi^{b} \eta_{ab}) &\sim \lambda_{2}
\sigma^{2} &; [\lambda_{2}] = &4 \\

\lambda_{3} J^{2} &\sim \lambda_{3} \sigma^{10} A^{8} &;
[\lambda_{3}] = &-4 \\

\lambda_{4} J &\sim \lambda_{4} \sigma^{5} A^{4} &; 
[\lambda_{4}] = &0 \\

\lambda_{5} \epsilon^{\alpha \beta \gamma \delta} \epsilon_{\alpha
\beta \gamma \delta} F^{a b}_{\alpha \beta} \nabla_{\lambda} \phi^{c}
\nabla_{\delta} \phi^{d} \phi^{e} &\sim \lambda_{5} \sigma^{3} A^{2}
\partial ^{2}~ {\rm or}~  \lambda_{5} \sigma^{3} A^{4} &;
[\lambda_{5}] = &0
\end{array} \)

\bigskip

Here $\sigma$ is taken to have mass dimension 0, $A$ to have mass
dimension 1, and the mass dimensions of the couplings are then as
indicated.  

In the last line the two terms arising from the spin
connection $A^{ab}_\alpha$ and vierbein $A^{5a}_\alpha$ pieces of $F$ are
indicated separately.  Also indicated is the occurrence of
derivatives, when one solves for the  $A^{ab}_\alpha$.   
The first of these terms corresponds to the usual Einstein action, and
leads us to identify the Planck mass according to 
\begin{equation}
M_{\rm Pl.}^2 \sim \lambda_5 \sigma^3 A^2 \label{eq:planckMass}.
\end{equation}
The second of these terms is an effective cosmological term.  Its
coefficient is given as
\begin{equation}
\lambda_5 \sigma^3 A^4 \sim M_{\rm Pl.}^4 / (\lambda_5 \sigma^3 ). 
\label{eq:MMcosTerm}
\end{equation} 
The terms in $J$ depend on the combination 
\begin{equation}
\sigma^5 A^4 \sim M_{\rm Pl.}^4/ (\lambda_5^2 \sigma).
\label{eq:Jcosterm}
\end{equation}
There is no $A$-dependence from the remaining terms, nor (we recall)
from conventional condensation energy.  

Thus if the pure numbers $\lambda_5, \sigma$ are large, the effective
cosmological term, appearing in the equations determining $A$, is
suppressed.  The largeness of $\lambda_5$ and $\sigma$ translates,
according to (\ref{eq:planckMass}), into the smallness of the mass scale
$A$.  This in turn points to a very large length scale associated to
the metric, such as might be induced by inflation.  In any case, it
seems significant that the effective cosmological term can here be
suppressed by a multiplicative, as opposed to a subtractive, mechanism.

\bigskip
\S8 Remarks 
\bigskip

\begin{enumerate}

\item If these considerations correspond to
reality, the structures assumed in general relativity, including the
conventional notion of distance and distinction of time, are not
primary.  Indeed, in the unbroken symmetry phase there is no
distinction among the variables describing space and time, and no
notion of distance (only volume).  It will be very interesting to
consider the possibility that a more symmetric phase was realized
in the early Universe, or near singularities of general relativity.

\item Our considerations have been entirely
classical.  The actions postulated here, although polynomial in the
fields, are of a very unfamiliar type for quantum field theory.
Considerations of a different order will be required to see if it is
possible to embed them into a consistent quantum theory.  This could
well require supersymmetry to soften the ultraviolet behavior, and
extra fields or extra dimensions
to cancel anomalies.  Since the symmetry-broken phase is
close to conventional general relativity, one might suspect that the
structures used in successful quantizations of that theory could also
be brought to bear here.  

\item There is no apparent reason why the
internal group could not be substantially larger.  It could break down
to a product of a non-trivial internal group and local Lorentz
invariance, rather than just the latter.  The occurrence, in
attractive unification schemes based on $SO(10)$, of representations
that are spinors both for the internal and the space-time symmetry is
suggestive in this regard.

\end{enumerate}

{\bf Acknowledgments} 
I wish to thank Mark Alford, John March-Russell and Lisa Randall for
valuable discussions.

\end{document}